\documentclass[notitlepage,a4paper,aps,prd,onecolumn,superscriptaddress,nofootinbib,groupedaddress]{revtex4}

\usepackage{enumerate}
\usepackage{amsmath}
\usepackage{amsfonts}
\usepackage{amssymb}
\usepackage[utf8]{inputenc}
\usepackage[T1]{fontenc}
\usepackage{mathtools}
\usepackage{wasysym}
\usepackage{accents}
\usepackage[svgnames]{xcolor}
\usepackage[colorlinks=true]{hyperref}
\usepackage{url}

\begin{document}
\title{Reaching the Planck scale with muon lifetime measurements}

\author{Iarley P. Lobo}
\email{lobofisica@gmail.com,  iarley_lobo@fisica.ufpb.br}
\affiliation{Department of Chemistry and Physics, Federal University of Para\'iba, Rodovia BR 079 - km 12, 58397-000 Areia-PB,  Brazil}
\affiliation{Physics Department, Federal University of Lavras, Caixa Postal 3037, 37200-900 Lavras-MG, Brazil}

\author{Christian Pfeifer}
\email{christian.pfeifer@zarm.uni-bremen.de}
\affiliation{ZARM, University of Bremen, 28359 Bremen, Germany.}
\affiliation{Laboratory of Theoretical Physics, Institute of Physics, University of Tartu, W. Ostwaldi 1, 50411 Tartu, Estonia.}

\begin{abstract}
Planck scale modified dispersion relations are one way how to capture the influence of quantum gravity on the propagation of fundamental point particles effectively. We derive the time dilation between an observer's or particle's proper time, given by a Finslerian length measure induced from a modified dispersion relation, and a reference laboratory time. To do so, the Finsler length measure for general first order perturbations of the general relativistic dispersion relation is constructed explicitly. From this we then derive the time dilation formula for the $\kappa$-Poincar\'e dispersion relation in several momentum space bases, as well as for modified dispersion relations considered in the context of string theory and loop quantum gravity. Most interestingly we find that the momentum Lorentz factor in the present and future colliders can, in principle, become large enough to constrain the Finsler realization of the $\kappa$-Poincar\'e dispersion relation in the bicrossproduct basis as well as a string theory inspired modified dispersion relation, at Planck scale sensitivity with the help of the muon's lifetime.
\end{abstract}

\maketitle

%%%%%%%%%%%%%%%%%%%%%%%%%%%%%%%%%%%%%%%%%%
\section{Introduction}
A major difficulty in the search for quantum gravity effects is that the scale at which they are expected to become relevant is at the Planck energy $E_{\text{P}}$ of order $10^{19}\, \text{GeV}$, respectively at distance of the Planck length $\ell_{\text{P}}$ of order $10^{-35}\, \text{m}$. Thus, in order to detect Planck scale effects one either needs to reach very high energies, or, probe very small lengths scales.

In the absence of a complete theory of quantum gravity, phenomenological models which shall capture aspects of quantum gravity, often employ Planck scale modified dispersion relations (MDRs) to effectively capture the interaction of particles, propagating through spacetime, with the quantum nature of gravity \cite{AmelinoCamelia:2008qg,Liberati:2013xla,Mattingly:2005re}. Such dispersion relations predict a deviation of particle trajectories from the general relativistic geodesics, with a leading order term in powers of the inverse Planck energy. Thus, MDRs lead to tiny corrections of the predictions of general relativity (GR), which are in principle detectable. To be able to detect these effects realistically, they need to be amplified, for example through accumulation over a long travel time of the particles. One observable, which meets this requirement and is accessible, is the time of arrival of high energetic gamma rays reaching us from gamma ray bursts at high redshift, for which Planck scale MDRs predict a dependence on the particles' energy \cite{Jacob:2008bw,Acciari:2020kpi,Pfeifer:2018pty,Rosati:2015pga,Amelino-Camelia:2016ohi}.

Recently, it was pointed out that in the comparison of lifetimes of particles and antiparticles (in particular for muons) Planck scale sensitivity for $\kappa$-deformations of the Poincar\'e algebra \cite{Lukierski:1991pn,Lukierski:1992dt,Majid:1994cy} is at reach \cite{Arzano:2019toz,Arzano:2020rzu}. In these considerations the \emph{momentum} Lorentz factor attained at particle accelerators plays the role of amplifier of the Planck scale effect.

Inspired by the promising findings of reaching Planck scale sensitivity with muons, we study the lifetime of elementary particles and the time dilation between their rest frame and a laboratory frame, induced by Planck scale MDRs. Following the famous clock postulate, that the proper time an observer measures between two events in its rest frame is the length of the observer's worldline between these events, the first ingredient necessary for our study is the length measure for wordlines induced by a MDR. In general, this will be a Finslerian length measure \cite{Pfeifer:2019wus,Girelli:2006fw,Raetzel:2010je,Amelino-Camelia:2014rga,Lobo:2016xzq}, i.e.\ a function $F(x,\dot x)$ depending on position and the velocity of an observer, which is $1$-homogeneous with respect to the $4$-velocity argument $\dot x$. Assuming a ``flat'' dispersion relation results in a flat length measure, i.e. independent of $x$, from which the derivation of the proper time along a worldline as function of the lab coordinate time can be done explicitly. 

We derive the time dilation formula between laboratory frame and particle rest frame for the most general MDR in first order perturbations of the GR dispersion relation explicitly, and apply the formula to several Planck scale MDRs motivated from the $\kappa$-Poincar\'e algebra, string theory and loop quantum gravity (LQG).

For the derivation we assume that the dispersion relation and the resulting length measure is universal for all massive physical objects, so in particular for the muon and the observer. On a curved spacetime, this ensures the implementation of the weak equivalence principle, i.e.\ that gravity couples universally to all physical objects, since the dispersion relation encodes the coupling between gravity and point particles (or the point particle limit of fields). Moreover, transformations which leave the dispersion relation invariant, then are candidates for observer transformations. For MDRs, these are modified Lorentz transformations. Hence, we are not analysing Lorentz invariance violating (LIV) scenarios, for which it is assumed that observers are still related by Lorentz transformations.

Throughout our calculations we carefully distinguish between the \emph{velocity Lorentz factor} $\gamma = \frac{1}{\sqrt{1-v^2}}$ and the \emph{momentum Lorentz factor} $\bar \gamma = \frac{p_0}{m}$, which in general do not coincide in the context MDRs.

Most interestingly, we explicitly calculate the time dilation of the lifetime of muons with energies available in facilities like the Large Hadron Collider (LHC), or the planned Future Circular Collider (FCC), at CERN, from the $\kappa$-Poincar\'e dispersion relation in the bicrossproduct basis (which is identically to the MDR derived from a D-brane string theory model). On the basis of a deformed relativity principle, which is implied by the Finsler geometric treatment of the modified dispersion relation\footnote{Although the string theory case is related to a Lorentz Invariance Violating (LIV) scenario, in contrast to the $\kappa$-Poincar\'e one that deforms Lorentz symmetry, we show that one can promote it to a deformed relativity scenario and analyse the resulting Muon lifetime phenomenology (an issue that is further discussed in section III).}, we find that the deformation parameter, $\kappa$ or $M_Q$, could be constrained by muon lifetime measurements at the colliders at the order of magnitude of the Planck energy, thus reaching Planck scale sensitivity for this quantum gravity phenomenology model with muon lifetimes.

In this article $\eta$ denotes the Minkowski metric $\textrm{diag}(+,-,-,-)$, indices $a,b,c,...$  run from $0$ to $3$ and indices $i,j,k,...$ run from $1$ to $3$. The symbol $\bar \partial^a = \frac{\partial}{\partial p_a}$ denotes derivative with respect to momentum coordinates.

%%%%%%%%%%%%%%%%%%%%%%%%%%%%%%%%%%%%%%%%%%%%%%%%%%%%%%%%%
\section{The time measure from general modified dispersion relations}
We briefly review the mathematical procedure to obtain a time measure from MDRs, before we apply the procedure to general first order modifications of the GR dispersion relation.

%%%%%%%%%%%%%%%%%%%%%%%%%%%%%%%%%%%%%%%%%%%%%%%%%%%%%%%%%
\subsection{The general algorithm}
Point particle dispersion relations are level sets of Hamilton functions $H(x,p)$ on the point particle phase space, technically the cotangent bundle $T^*M$ of spacetime. To associate a time measure to massive point particles, resp. observers, from the dispersion relation one employs the Helmholtz action of free particles \cite{Girelli:2006fw,Raetzel:2010je,Pfeifer:2019wus,Amelino-Camelia:2014rga,Letizia:2016lew,Lobo:2016xzq,Lobo:2016lxm}
\begin{align}\label{eq:actHelm}
    S[x,p,\lambda]_{H} = \int d\mu (\dot x^a p_a - \lambda f(H(x,p),m))\,,
\end{align}
where $\mu$ is an arbitrary curve parameter, ``dot'' means derivative with respect to this parameter, $f$ is a function such that $f=0$ is equivalent to the dispersion relation $H(x,p) = m^2$ and $\lambda$ is a Lagrange multiplier.

To obtain a length measure for massive particle trajectories from this action we use the following algorithm:
\begin{enumerate}
    \item Variation with respect to $\lambda$ enforces the dispersion relation.
    \item Variation with respect to $p_a$ yields an equation $\dot x^a = \dot x^a(p,\lambda)$ which must be inverted to obtain $p_a(x,\dot x, \lambda)$ to eliminate the momenta  $p_a$ from the action.
    \item Using $p_a(x,\dot x, \lambda)$ in the dispersion relation, one can solve for $\lambda(x,\dot x)$.
    \item Finally the desired length measure is obtained as $S[x] =  S[x,p(x,\dot x,\lambda(x,\dot x)),\lambda(x,\dot x)]_{H}$.
\end{enumerate}
The crucial step in this algorithm is to be able to find $p_a(x,\dot x, \lambda)$, i.e.\ to invert the relation $\dot x^a = \dot x^a(p,\lambda)$. If this is globally possible, only locally, or not at all, depends on the dispersion relation under consideration and the choice of the function $f$ \cite{Raetzel:2010je,Rockafellar}. Among others, choices employed in the literature are $f(H,m) = \ln(H(x,\frac{p}{m}))$, for homogeneous Hamiltonians \cite{Raetzel:2010je}, or, $f = H(x,p) - m^n$, often with $n=2$ \cite{Amelino-Camelia:2014rga,Letizia:2016lew}, for nonhomogeneous Hamiltonians. Another suggestions for the inhomogeneous case is $f(H,m) = \ln(\frac{H(x,p)}{m^n})$, which is not yet investigated in detail.

%%%%%%%%%%%%%%%%%%%%%%%%%%%%%%%%%%%%%%%%%%%%%%%%%%%%%%%%%
\subsection{First order modified dispersion relations}
Let $g$ be a general Lorentzian metric and $h(x,p)$ be a function on the cotangent bundle $T^*M$ of spacetime. A general first order modification of the GR dispersion relation is defined by the Hamilton function
\begin{align}\label{eq:Ham}
    H(x,p) = g(p,p) + \epsilon h(x,p)\,.
\end{align}
The term $g(p,p) = g^{ab}(x)p_a p_b$ defines the GR point particle dispersion relation on a curved spacetime, $\epsilon$ is a perturbation parameter, counting the first nontrivial correction to GR, and $h(x,p)$ a perturbation function, which needs to be specified depending on the application in consideration. In the context of quantum gravity phenomenology, $\epsilon$ is usually related to the Planck length or Planck energy and $h(x,p)$ can, for example, be obtained from Planck scale MDRs, such as the $\kappa$-Poincar\'e dispersion relation and others, whose influence we investigate in the course of this paper.

The steps of the previously outlined algorithm can now be performed as follows:
\begin{enumerate}
    \item Variation of the action \eqref{eq:actHelm} with respect to $\lambda$ yields $f=0$, which in turn enforces the dispersion relation
    \begin{align}\label{eq:mdr}
        g(p,p) + \epsilon h(x,p) = m^2\, .
    \end{align}
    \item Variation of the action \eqref{eq:actHelm} with respect to $p_a$ and using the perturbative Hamiltonian \eqref{eq:Ham} yields
    \begin{align}
        \dot x^a 
        = \lambda \bar\partial^a H\partial_H f 
        = \lambda (2 p^a + \epsilon \bar\partial^a h)\partial_H f\,,
    \end{align}
    which can be rewritten as (indices are raised on lowered with the components of the metric $g$)
    \begin{align}\label{eq:p}
        p_a = \frac{\dot x_a}{2 \lambda \partial_Hf} - \frac{1}{2}\epsilon g_{ab} \bar\partial^bh\,.
    \end{align}
    Explicitly inverting this equation completely to obtain $p_a(x,\dot x,\lambda)$ is not possible, but also not necessary in perturbation theory, as we will see soon. We note the following two relevant relations:
    \begin{align}
        \dot x^a p_a 
        &= \lambda ( 2 g(p,p)  + \epsilon p_a \bar\partial^a h) \partial_Hf\,,\label{eq:dxp}\\
        g(p,p)
        &= \frac{g(\dot x, \dot x)}{4 \lambda^2 \partial_Hf^2} - \epsilon \frac{\dot x_a \bar\partial^a h}{2 \lambda \partial_Hf}\, ,\label{eq:gpp}
    \end{align}
    where we introduced the notation $g(\dot x, \dot x)=g_{ab}(x)\dot x^a \dot x^b$.
    \item Using \eqref{eq:gpp} in the dispersion relation \eqref{eq:mdr}, using a first order expansion of the Lagrange multiplier $\lambda = \lambda_0 + \epsilon \lambda_1$ and solving the dispersion relation order by order leads to
    \begin{align}\label{eq:l}
        \lambda_0 = \frac{\sqrt{g(\dot x,\dot x)}}{2 m \partial_Hf}\,,\quad
        \lambda_1 = \frac{\sqrt{g(\dot x, \dot x)}}{4 m^3 \partial_Hf}h - \frac{\dot x_a \bar\partial^a h}{4 m^2 \partial_Hf}\,.
    \end{align}
    \item Combining all the results from \eqref{eq:dxp}, \eqref{eq:gpp} and \eqref{eq:l} in \eqref{eq:actHelm} for the Hamiltonian \eqref{eq:mdr} the action becomes
    \begin{align}\label{eq:S}
        S[x] = \int d\tau \sqrt{g(\dot x, \dot x)}\left(m -\epsilon \frac{h}{2m} \right).
    \end{align}
    At this order, the function $f$ and its derivatives all cancel and so the specific choice is not relevant. The perturbation function $h$ appearing in \eqref{eq:S} has to be understood as $h=h(x,\bar p(x,\dot x))$, with $\bar p_{a}(x,\dot x) = m\frac{\dot x_a}{\sqrt{g(\dot x, \dot x)}}$.
\end{enumerate}

We have proven that the Finsler function which governs the massive point particle motion of first order MDR is
\begin{align}\label{eq:FinsOrig}
    F(x,\dot x) = m \sqrt{g(\dot x, \dot x)} - \epsilon \sqrt{g(\dot x, \dot x)} \frac{h(x,\bar p(x,\dot x))}{2m}\,.
\end{align}
As an example, consider an $n$-th order polynomial modification
\begin{align}
    h(x,p) = h^{a_1 a_2 ....a_n}(x)p_{a_1} p_{a_2}...p_{a_n} \\
    \Rightarrow h(x,\bar p(x,\dot x)) =  m^n \frac{h_{a_1 a_2 ....a_n}(x)\dot x^{a_1} \dot x^{a_2} ...\dot x^{a_n}}{g(\dot x,\dot x)^\frac{n}{2}}
\end{align}
which yields
\begin{align}\label{F1}
     F(x,\dot x) = m \sqrt{g(\dot x, \dot x)} - \epsilon m^{n-1} \frac{h_{a_1 a_2 ....a_n}(x)\dot x^{a_1} \dot x^{a_2} ...\dot x^{a_n}}{2g( \dot x,\dot x)^{\frac{n-1}{2}}}\,.
\end{align}

%%%%%%%%%%%%%%%%%%%%%%%%%%%%%%%%%%%%%%%%%%%%%%%%%%%%%%%%%
\section{The Muon lifetime from modified dispersion relations}\label{sec:mlife}

Next, we analyze how the lifetime of a fundamental particle is modified by the assumption that it propagates on a Finsler spacetime, \cite{Hohmann:2018rpp,Javaloyes:2018lex}, induced by a MDR. The clock postulate is implemented in the following way.

The proper time an observer, or massive particle, experiences  between events $A$ and $B$ along a timelike curve (her worldline) in a Finsler spacetime $({\cal M},F)$ is the length of this curve between events $A$ and $B$:
\begin{equation}\label{ptime}
    \Delta\tau_{AB}\doteq m^{-1}\int_{\mu_A}^{\mu_B}F(x,\dot{x})d\mu\,.
\end{equation}

We aim to investigate the decay of fundamental particles in accelerators, therefore we shall discard pure gravitational effects, i.e, the spacetime curvature, and rely on Finsler-deformations of Minkowski proper time. Mathematically this is justified by the existence of special coordinates, which allow one to neglect curvature effects at small coordinate distance around every point and a given direction on Finsler spacetimes \cite{Pfeifer:2014eva}. Thus to zeroth order we consider $g(\dot{x},\dot{x})$ in Eq.(\ref{F1}) as the usual Minkowski metric, which we shall label $\eta(\dot{x},\dot{x})$. In Cartesian coordinates we simply write
\begin{align}
\eta(\dot{x},\dot{x})= (\dot{x}^0)^2-\delta_{ij} (\dot{x}^i)(\dot{x}^j)\,.
\end{align}

Since the arc length is invariant under reparametrizations, we transform the arbitrary parameter $\mu$ to the time coordinate in the laboratory frame, $x^0\doteq t$ in \eqref{ptime}. Using \eqref{F1}, we have the following modification of the proper time between the events with parameters $(x^0)_A=t_A$ to $(x^0)_B=t_B$ (from now on we omit the label ``$AB$'' in $\Delta\tau_{AB}$):
\begin{align}\label{eq:mdrfins}
    \Delta\tau=\int_{t_A}^{t_B} dt \left[ \gamma^{-1} - \frac{\epsilon}{2} m^{n-2} \gamma^{n-1} h_{a_1...a_n}\frac{dx^{a_1}}{dt}...\frac{dx^{a_n}}{dt}\right]\,,
\end{align}
where, for convenience, we introduced the ususal \emph{velocity Lorentz factor}
\begin{equation}\label{eq:gamma}
    \gamma=\frac{1}{\sqrt{1-v^2}}\, ,
\end{equation}
with $v^i\doteq dx^i/dt$ and $v^2 = \delta_{ij}v^i v^j$.

Suppose a fundamental particle has its lifetime dilated in a circular accelerator, like the LHC or the FCC \cite{fcc}. In this case, the norm of the three-dimensional velocity $v^2$ is roughly a constant, which allows for a simplification of the above expression. In the following, $\Delta t$ will be the time measured in the laboratory frame in which the particle is accelerated, while $\Delta\tau$ is the proper time experienced by the particle, respectively measured by an observer comoving to the particle.

%%%%%%%%%%%%%%%%%%%%%%%%%%%%%%%%%%%%%%%%%%%
\subsection{The $\kappa$-Poincar\'e dispersion relation in bicrossproduct basis type}\label{kappa}

For the $\kappa$-Poincar\'e dispersion relation in the bicrossproduct basis, the first order correction of the quadratic GR Finsler function is a polynomial of degree $n=3$. The symbols $h_{a_1a_2a_3}$ for this case are $h_{a_1a_2a_3} = -\frac{1}{3}(\delta^0_{a_1} \delta_{ij}\delta^i_{a_2}\delta^j_{a_3} + \delta^0_{a_2} \delta_{ij}\delta^i_{a_1}\delta^j_{a_3} + \delta^0_{a_3} \delta_{ij}\delta^i_{a_1}\delta^j_{a_2} )$ and null otherwise, see for example \cite{Gubitosi:2013rna}. Hence the integrand in \eqref{eq:mdrfins} becomes
\begin{equation}\label{integrand}
    \gamma^{-1} + \epsilon \frac{m}{2} \gamma^{2} v^2 = \gamma^{-1}\left(1 + \epsilon \frac{m}{2} \gamma(\gamma^2-1)\right)\,.
\end{equation}
Moreover, $\epsilon$ is a parameter expected to be of the order of the inverse of the energy scale at which quantum gravitational corrections are expected to take place, which we simply denote as deformation parameter $\kappa^{-1}$, as it is usually done in the context of the $\kappa$-Poincar\'e algebra.

Therefore, we express the lifetime of a fundamental particle that probes a Finsler spacetime induced by the $\kappa$-Poincar\'e dispersion relation in the bicrossproduct basis as (we define $\Delta t\doteq t_B-t_A$)
\begin{align}
    \Delta\tau=\frac{\Delta t}{\gamma}\left[1+\frac{m}{2\kappa}\gamma(\gamma^2-1)\right].
\end{align}
This geometric invariant quantity defined by Eq. \eqref{ptime} measures the proper time a particle experiences, and is related to the time which passes in the laboratory, with respect to which the particle is accelerated. Thus the measured lifetime of a particle in a laboratory, denoted by $\Delta t$, can be related to the proper lifetime of the particle $\Delta \tau$, depending on its coordinate velocity $v$ through the factor $\gamma$. To first order $\kappa^{-1}$, we find for the laboratory frame lifetime of the particle
\begin{align}\label{eq:DTgamma}
\Delta t= \gamma \Delta \tau \left[1-\frac{m}{2\kappa}\gamma(\gamma^2-1)\right]\, .
\end{align}

In order to compare with data from particle accelerators, we need to express the velocity $\gamma$ factor defined in \eqref{eq:gamma} in terms of the energy $p_0$ and mass $m$ of the particles. The conversion between these dependencies is nontrivial due to the deviations from the usual relativistic setting. We derive the $4$-momentum of the particles, which satisfies the MDR:
%~\footnote{The first equality below can be verified by direct calculation. Using the Lagrange multiplier \eqref{eq:l} in \eqref{eq:p}, to express $p_a$ as function of $\dot x$, coincides with $p_a = \frac{\partial}{\partial \dot x^a}F$ when using $F$ as identified in \eqref{eq:FinsOrig}.}:
\begin{align}
    p_0 
    = \frac{\partial}{\partial \dot x^0} F(x,\dot x) 
    = m \gamma  - \frac{m^2}{2\kappa}(\gamma^2 - 1)(2\gamma^2 - 1)\,,\quad
    p_i 
    = \frac{\partial}{\partial \dot x^i} F(x,\dot x) 
    = m \gamma v_i \left( -1 + \frac{m}{\kappa} \gamma (2 - \gamma^2) \right)\,.  
\end{align}
Solving the first relation for $\gamma$ as a function of $p_0$ yields $\gamma = \frac{p_0}{m} + \frac{m}{2\kappa}\left(1 - 3 \frac{p_0^2}{m^2} + 2 \frac{p_0^4}{m^4} \right)$. Employing this in \eqref{eq:DTgamma} gives us the lifetime as a function of $p_0$
\begin{align}\label{eq:kappaDT}
    \Delta t
    =\Delta t_{\text{SR}} \left[1 + \frac{m}{2\kappa} \left( \frac{m}{p_0} - 2 \left(\frac{p_0}{m}\right) +  \left(\frac{p_0}{m}\right)^3 \right)\right]\,,
\end{align}
where $\Delta t_{\text{SR}}=\frac{p_0}{m}\Delta\tau$ is the usual special relativistic dilated lifetime expressed in terms of the particle's $p_0$ component. This result leads us to introduce the \emph{momentum Lorentz factor} $\bar \gamma = \frac{p_0}{m}$. We would like to emphasize that for MDRs, in general, the momentum Lorentz factor is different from the velocity Lorentz factor, as we have demonstrated by the derivation of the relation $\gamma(\bar\gamma) = \bar\gamma + \frac{m}{2\kappa}\left(1 - 3 \bar\gamma^2 + 2 \bar\gamma^4 \right)$. 

Before we continue we would like to point out a short general comment on deformed Lorentz transformations. In general, the $4$-momentum defined as $p_a = \frac{\partial}{\partial \dot x^a}F$ satisfies the deformed dispersion relation. This can be proven by using the Lagrange multiplier \eqref{eq:l} in \eqref{eq:p}, to express $p_a$ as a function of $\dot x$, which coincides with $p_a = \frac{\partial}{\partial \dot x^a}F$ when using $F$ as identified in \eqref{eq:FinsOrig}. In fact, for a particle at rest ($v=0,\gamma=1$), the dispersion relation implies for the momenta $p_0 = m$ and $p_i = 0, i=1,2,3$. If we consider the momenta as function of $\gamma$ and apply the transformation 
\begin{align}\label{eq:pofgamma}
    p_0=p_0(1) \to p_0' = p_0(\gamma) = \frac{\partial}{\partial \dot x^0}F \textrm{ and } p_i=p_i(1)\to p_i'= p_i(\gamma) = \frac{\partial}{\partial \dot x^i}F\,,
\end{align}
then the dispersion relation $H(x,p) = H(x,p') = m^2$ is satisfied. Therefore, the transformation that links the $4$-momentum of the particle at rest to the $4$-momentum $\left(p_0(\gamma),p_i(\gamma)\right)$ represents a deformed Lorentz transformation.

From \eqref{eq:kappaDT}, we are able to identify the dimensionless quantity $\delta_{p_0,m}$, depending on the mass and energy of the particles attained in accelerators, which is responsible for an effect beyond special relativity and is the one which we compare with the uncertainty of the most precise experimental values of the mean lifetime of fundamental particles:
\begin{align}\label{delta1}
   \delta_{p_0,m} = \frac{m}{2\kappa} \left( \bar \gamma^{-1} - 2 \bar \gamma + \bar \gamma^3 \right) \approx \frac{m}{2 \kappa}\bar \gamma^3\,.
\end{align}
In the last approximation we focused on the term which dominates for high energetic particles. 

For a concrete example, let us consider the case of the muon particle. The muon mean lifetime amounts to \cite{Zyla:2020zbs}
\begin{equation}\label{ptmuon}
    \tau_{\mu}=(2.1969811 \pm 0.0000022) \times 10^{-6}\, \text{s}=2.1969811\, \mu\text{s}\pm \sigma_{\tau},
\end{equation}
and its most precise measurement was done for low energy muons in \cite{Tishchenko:2012ie}. From (\ref{ptmuon}), we see that the relative uncertainty of this measurement reads
\begin{align}\label{sigma}
\sigma_{\tau}/\tau_{\mu}\approx 10^{-6}.
\end{align}

In the following, we shall explore the consequences of assuming that experiments in the LHC or the FCC could measure the muon lifetime with the same relative uncertainty, which, as we shall demonstrate, would allow one to set significant constraints on the quantum gravity energy scale. An analogous prediction was done previously for the case of the decays of the muon and antimuon, for a $\kappa$-Poincar\'e basis in which the Hamiltonian is undeformed, and modifications take place when comparing the lifetimes of particles and antiparticles in the context of CPT violation~\cite{Arzano:2019toz,Arzano:2020rzu}. 

As a matter of fact, had we used the same basis of \cite{Arzano:2019toz}, i.e., with an undeformed Casimir operator as Hamilton function, we would have derived the standard Minkowski metric, without Finsler modifications, thus producing no effect beyond special relativity in the lifetime of particles depending on their relative velocity. We should stress that this is a general feature of the use of different coordinates in curved momentum spaces, i.e., different momentum space bases lead to inequivalent relativistic theories and predictions \cite{Amelino-Camelia:2019dfl}, see also when we discuss isotropic dispersion relations in the next section. As we shall see now, we will be able to increase the estimated bound from lifetime observations in 2 orders of magnitude in comparison to previous approaches~\cite{Arzano:2020rzu}. 

Comparing \eqref{delta1} and \eqref{sigma}, we can estimate a lower bound for the $\kappa$ parameter using the momentum Lorentz factor,~$\bar{\gamma}$, achieved in facilities like the LHC ($p_0/m\sim 10^4$) or that shall be achieved in the FCC ($p_0/m\sim 10^5$) \cite{Arzano:2020rzu}\footnote{These estimates are based on energies $6.5\, \text{TeV}$ for the LHC \cite{Aaboud:2016mmw}, and $50\, \text{TeV}$ for the FCC \cite{Benedikt:2018csr}.}. Using the mass of the muon \cite{Zyla:2020zbs}
\begin{equation}
    m_{\mu}\approx 105.6583745\, \text{MeV}
\end{equation}
and $\bar \gamma_{\text{LHC}} = 10^4$ we find the LHC upper bound as
\begin{align}\label{eq:kappabound1}
    \kappa_{\text{LHC}}\geq \frac{m_\mu \bar{\gamma}^3_{\text{LHC}}}{2} \frac{\tau_\mu}{\sigma_\tau}\approx 5.3\times 10^{16}\, \text{GeV},
\end{align}
which lies 3 orders of magnitude below the Planck energy $E_{\text{P}}\approx 1.22\times 10^{19}\, \text{GeV}$ and corresponds to the scale of some inflationary models \cite{Tegmark:2004qd}. This is already an interesting result, since it is 2 orders of magnitude higher than the bound proposed in \cite{Arzano:2019toz,Arzano:2020rzu}.

Using the optimal $\bar{\gamma}$ factor which can be reached by the LHC for muons from $\bar{\gamma}_{\text{LHCopt}} = 6.5 \text{TeV} /m_{\mu} = 6.1 \times 10^4$ one even reaches
\begin{align}\label{eq:kappabound2}
    \kappa_{\text{LHCopt}}\geq \frac{m_\mu \bar{\gamma}_{\text{LHCopt}}^3}{2} \frac{\tau_\mu}{\sigma_\tau}\approx 1.2 \times 10^{19}\, \text{GeV} \sim E_{\text{P}}\,.
\end{align}

We should stress that the assumption of reaching these optimal conditions, like the precision (\ref{sigma}) in the LHC, are maximally optimistic. For instance, the measurement of short-lived hadrons has been recently performed at the CMS with relative uncertainty of order ${\cal O}(10^{-2})$ \cite{CMS-PAS-BPH-13-008}. Some extra difficulties arise when using the muon decay due to its very long lifetime \footnote{Nevertheless, the deformed muon lifetime is a good candidate effect due to the smallness of the muon's mass, which works as an amplifier.}. However, the decay of exotic long-lived particles (with lifetimes of the order $10\, \text{ps}$ to $10\, \text{ns}$) also have been searched in some LHC experiments \cite{Lee:2018pag}, and there is room for improvement in this analysis \cite{Banerjee:2019ktv}.

However, latest with the next generation colliders, such as the FCC, the muon lifetime shall be amplified by the Lorentz factor $\bar{\gamma}_{\text{FCC}}=4.7\times 10^5$, which alleviates the needed relative uncertainty to constrain this effect at the Planck scale: it could be of the order ${\cal O}(10^{-4})$ to ${\cal O}(10^{-3})$, which lies close to current capabilities. Besides that, this observable represents an unforeseen opportunity for testing Planck scale physics in prospective facilities like Muon Colliders \cite{Delahaye:2019omf}, where the dilated lifetime could be measured in longer baselines at the TeV scale.

Before moving on to further MDRs we point out that the dispersion relation of the type $H =  \eta(p,p) + \epsilon p_0 \delta^{ij} p_i p_j$ does not only emerge in the context of the $\kappa$-Poincar\'e algebra, but also in the context of propagation of particles in a quantum spacetime modeled by D-brane fluctuations in string theory \cite{Ellis:1999uh}. In our approach, the LIV nature of this string theory dispersion relation gets supplemented by deformed Lorentz transformations induced by \eqref{eq:pofgamma}. The bounds \eqref{eq:kappabound1} and \eqref{eq:kappabound2}, then translate into a bound on the quantum gravity scale $\frac{\xi}{M_{\text{QG}}}$, for this LIV to DSR lifted model, which is obtained by replacing $\kappa$ by $\frac{M_{\text{QG}}}{2\xi}$. The bounds do not apply to the original LIV string theory model.

%%%%%%%%%%%%%%%%%%%%%%%%%%%%%%%%%%%%%%%%%%%%%%%%%%%%%%%
\subsection{Isotropic modified dispersion relations}\label{others}
We extend our analysis of time dilations to general MDRs which are rotational invariant, i.e.\ depend only on the norm of the spatial momentum $q=\sqrt{\delta^{ij}p_i p_j}$. The time measuring Finsler function \eqref{eq:FinsOrig} becomes
\begin{align}
    F(\dot x) = m \sqrt{\eta(\dot x, \dot x)}\left( 1 - \epsilon \frac{h(\bar p_0(\dot x),\bar q(\dot x))}{2 m^2}\right) \,.
\end{align}
Using the relation $\bar p_a = m \frac{\dot x_a}{\sqrt{\eta(\dot x, \dot x)}}$, the reparametrization invariance of the time measure \eqref{ptime} and the notation from the previous section for $\bar p_0(\dot x) = m \frac{\dot x^0}{\sqrt{\eta(\dot x, \dot x)}} = m \gamma$ and $q=\sqrt{\delta^{ij}\bar p_i(\dot x) \bar p_j(\dot x)} = m \sqrt{\tfrac{\delta_{ij}\dot x^i \dot x^j}{\eta(\dot x,\dot x)}} = m \gamma v$,  we obtain the general time dilation formula for this kind of MDRs
\begin{align}
   \Delta t = \gamma \Delta \tau \left[ 1 + \frac{\epsilon}{2 m^2} h(m \gamma, m \sqrt{\gamma^2-1}) \right]\,,
\end{align}
where the units of the leading order perturbation parameter $\epsilon$ must be adopted depending on the choice of $h$. Often the leading order terms beyond special relativity are characterized by a polynomial $h = \sum_{r,s} \sigma_{rs} p_0^r q^s$, where $\sigma_{rs}$ are numerical coefficients and $r,s$ are integers. For these modifications the time dilation in terms of the velocity Lorentz factor becomes 
\begin{align}\label{dtgeneral}
    \Delta t = \gamma \Delta \tau\left[ 1 +  \frac{1}{2}\sum_{r,s} \sigma_{rs}  \left(\frac{m}{E_{\text{P}}}\right)^{r+s-2} \gamma^{r} (\gamma^2-1)^{\frac{s}{2}} \right]\,.
\end{align}
To compare this dilation formula directly with the lifetime of particles of a certain energy, one needs to rewrite this expression in terms of the momentum Lorentz factor $\bar\gamma$. Therefore it is necessary to derive the relation between $\gamma$ and $(p_0,m)$ case by case, analogously as we presented in the previous section before \eqref{eq:kappaDT}.

To conclude, we list several prominent modification functions $h$ and their particle lifetime prediction in terms of the velocity Lorentz factor $\gamma$ in Table \ref{tab:1}
\begin{table*}[ht]
    \centering
	\begin{tabular}{|l|l|l|}
		\hline
		Type and Theory & Perturbation function & Time dilation \\ \hline
		\begin{tabular}
		{@{}l@{}} {\textbf{Monomials}} \\ $r=1,s=2$: D-brane recoil \cite{Ellis:1999sd} \\ $r=1,s=2$: $\kappa$-Poincar\'e bicrossproduct \cite{Gubitosi:2013rna}, D-brane foam \cite{Ellis:1999uh} \\ $r=0,s=3$: example from \cite{Girelli:2006fw}, Liouville-String QG \cite{AmelinoCamelia:1996pj} \\ $r=4,s=0$: LQG inspired MDRs \cite{Amelino-Camelia:2016gfx,Brahma:2018rrg,Lobo:2019jdz}
		\end{tabular} 
		& $h = \alpha p_0^r q^s$ &
		\begin{tabular}
		{@{}l@{}}  $\Delta t = \gamma \Delta\tau  \left[ 1 + \frac{\alpha}{2}\left(\frac{m}{E_{\text{P}}}\right)^{w} \gamma^{r} (\gamma^2-1)^{\frac{s}{2}} \right]$ \\ \hspace{55pt} $w=r+s-2$
		\end{tabular}
		\\ \hline
		\begin{tabular}
		{@{}l@{}} \textbf{$\mathbf{p_0}$ powers} \\ $r=1$: $\kappa$-Poincar\'e Magueijo-Smolin basis \cite{KowalskiGlikman:2002we} 
		\end{tabular} 
		& $h= \alpha (p_0^2 - q^2) p_0^r $ & $\Delta t = \gamma \Delta\tau \left[ 1 + \frac{\alpha}{2} \left(\frac{m}{E_{\text{P}}}\right)^r \gamma^r \right]$ \\ \hline
		\begin{tabular}{@{}l@{}} 
		\textbf{Metric factor powers} \\ $s=2$: $\kappa$-Poincar\'e preferred basis in \cite{Relancio:2020rys} 
		\end{tabular} 
		& $h=\alpha (p_0^2 - q^2)^s$ & $\Delta t = \gamma \Delta\tau \left[ 1 +\frac{\alpha}{2} \left(\frac{m}{E_{\text{P}}}\right)^{2(s-1)} \right]$ \\ \hline
	\end{tabular}
	\caption{Time dilation formulas for different MDRs.}
\label{tab:1}
\end{table*}

With our findings we added another piece to the systematic analysis of Planck scale MDRs and their predictions of observables. Surprisingly, for first order in Plank energy corrections, the Planck scale sensitivity for muon lifetimes lies in reach already with the~LHC, under optimal conditions, but latest with the planned~FCC for more attainable requirements.

%%%%%%%%%%%%%%%%%%%%%%%%%%%%%%%%%%%%%%%%%%%%%%%%%%%%%%%
\section*{Acknowledgments}
The authors thank Nick E. Mavromatos, Albert de Roeck, and Alice Florent for the insightful comments. C.P. was supported by the Estonian Ministry for Education and Science through the Personal Research Funding Grant PSG489, as well as the European Regional Development Fund through the Center of Excellence TK133 ``The Dark Side of the Universe'' and was funded by the Deutsche Forschungsgemeinschaft (DFG, German Research Foundation) - Project Number 420243324. I.P.L. was partially supported by the National Council for Scientific and Technological Development - CNPq grant 306414/2020-1. The authors would like to acknowledge networking support by the COST Action QGMM (CA18108), supported by COST (European Cooperation in Science and Technology).

%%%%%%%%%%%%%%%%%%%%%%%%%%%%%%%%%%%%%%%%%%
%%%%%%%%%%%%%%%%%%%%%%%%%%%%%%%%%%%%%%%%%%

%\appendix

%%%%%%%%%%%%%%%%%%%%%%%%%%%%%%%%%%%%%%%%%%
%\section{Appx}

%%%%%%%%%%%%%%%%%%%%%%%%%%%%%%%%%%%%%%%%%%%%%%%%%%%%%%%
%%%%%%%%%%%%%%%%%%%%%%%%%%%%%%%%%%%%%%%%%%%%%%%%%%%%%%%
\bibliographystyle{utphys}
\bibliography{MDRMuon}

\end{document}